\documentclass[fleqn,12pt]{wlscirep}
\usepackage[utf8]{inputenc}
\usepackage[T1]{fontenc}
\usepackage[numbers,super]{natbib}
\usepackage{caption}

\newcommand{\farcs}{\mbox{\ensuremath{.\!\!^{\prime\prime}}}}%  % fractional arcsecond symbol: 0.''0
\newcommand{\apj}{Astrophys. J.}   % Astrophysical Journal
\newcommand{\apjl}{Astrophys. J. Lett.}   % Astrophysical Journal, Letters
\newcommand{\apjs}{Astrophys. J. Suppl. Ser.}   % Astrophysical Journal, Supplement
\newcommand{\aap}{Astron. Astrophys.}   % Astronomy and Astrophysics
\newcommand{\aaps}{Astron. Astrophys. Suppl.}   % Astronomy and Astrophysics, Supplement
\newcommand{\mnras}{Mon. Not. R. Astron. Soc.}   % Monthly Notices of the RAS

\newcommand{\nat}{Nature} % Nature

\newcommand{\nastro}{Nat. Astron.} % Nature Astronomy
\newcommand{\pasp}{Publ. Astron. Soc. Pacific}   % Publications of the Astron. Soc. of the Pacific
\graphicspath{{./}{figures/}}

\title{AGN ruled out as the dominant source of cosmic reionization}

\author[1,2]{Danyang Jiang}
\author[1,2,*]{Linhua Jiang}
\author[1,2]{Shengxiu Sun}
\author[1,2]{Weiyang Liu}
\author[1,2]{Shuqi Fu}

\affil[1]{Department of Astronomy, School of Physics, Peking University, Beijing 100871, China}
\affil[2]{Kavli Institute for Astronomy and Astrophysics, Peking University, Beijing 100871, China}
\affil[*]{e-mail: jiangKIAA@pku.edu.cn}

\makeatletter
\renewcommand{\@maketitle}{%
{%
\thispagestyle{empty}%
\vskip-36pt%
{\raggedright\sffamily\bfseries\fontsize{20}{25}\selectfont \@title\par}%
\vskip10pt
{\raggedright\sffamily\fontsize{12}{16}\selectfont  \@author\par}
\vskip25pt%
}%
}%
\makeatother

\begin{document}

\flushbottom
\maketitle

%\noindent $^*$corresponding authors\\
%\noindent $^\dagger$All author affiliations are listed at the end of the paper\\

\noindent 
\textbf{Cosmic reionization represents the latest phase transition in the Universe, when the Lyman continuum (LyC) photons turned the intergalactic medium (IGM) from neutral to highly ionized. It has long been debated whether galaxies or active galactic nuclei (AGNs) are the major source of LyC photons responsible for reionization. Previous observations slightly favored galaxies as the major ionizing source. However, the James Webb Space Telescope (JWST) recently discovered an unexpectedly high density of AGN candidates at high redshift, which has largely enhanced the influence of AGNs. Here we derive a definitive upper bound on the AGN contribution to reionization using the latest JWST data, and conclusively rule out AGNs as the dominant ionizing source during the peak epoch of reionization (EoR). We build a sample of galaxies and AGNs in a specific redshift range $7.15 \leq z \leq 7.75$ with a high completeness. Each object is then decomposed into a point-source component and an extended component in their rest-frame far-UV JWST images. We assume all point-source components are AGNs. Our fiducial AGN sample reaches an unprecedentedly low luminosity of $M_{\rm UV} \approx -15$ mag. Based on this sample, AGNs can contribute at most one third of the LyC photons budget required at $z\sim7.5$. Our result implies that galaxies dominate the ionizing source during the EoR.}

%%Introduction
The major source of LyC photons responsible for cosmic reionization remains controversial, because the individual contributions of the two main sources, galaxies (or stars) and AGNs (or supermassive black holes), are not well constrained \cite{miralda-escude2000,volonteri2009,onoue2017}. JWST is revolutionizing our understanding of reionization by routinely finding high-redshift ($z\ge6$) galaxies and AGNs. It is found that galaxies can produce sufficient LyC photons if their escape fraction $f_{\rm esc}$ is higher than $10\sim15$\% (ref.\cite{mascia2024,whitler2024}). However, the average $f_{\rm esc}$ value of lower-redshift galaxies is only a few percent \cite{grazian2017,liu2023}, and it is unrealistic to directly detect the LyC emission from high-redshift objects due to the IGM absorption. On the other hand, the UV luminosity function (LF) of luminous AGNs (i.e., quasars) has been well established \cite{matsuoka2018,schindler2023}. Despite that the faint end of the AGN LF is very uncertain, previous studies suggest that the AGN contribution to reionization is small \cite{matsuoka2018,jiang2022,schindler2023}. Recently, JWST discovered a large number of faint AGN candidates at $z\ge4$ (ref.\cite{onoue2023,harikane2023,kocevski2023,kocevski2024,kokorev2024,matthee2024,greene2024,williams2024}), including the so-called `little-red-dots' (LRDs) whose nature is still unclear \cite{li2024,zhang2024,labbe2024}. If the majority of these objects are AGNs, the AGN spatial density at the faint end would be hundreds of times higher than the extrapolated value from the current AGN LF \cite{harikane2023,kocevski2024,kokorev2024,matthee2024}. This leads to a possibility that AGNs play an important role on producing ionizing photons\cite{madau2024}. Meanwhile there are evidences supporting a subdominant contribution from AGNs \cite{Padmanabhan2023,Dayal2025}.

%Methodology
In this work we derive the absolute upper bound of the AGN contribution to reionization without knowing the nature of LRDs (and galaxies in general). We first build a sample of objects at $7.15 \leq z \leq 7.75$, including galaxies and AGNs, using JWST images in the GOODS (GOODS-S and GOODS-N) fields. This redshift range covers the peak of cosmic reionization\cite{planck2020}. It is optimal for our target selection, because the strong Ly$\alpha$ break caused by the IGM absorption produces very red $\rm F090W-F115W$ colors, which is highly efficient to remove low-redshift contaminants. Each of the selected object is then decomposed into a point-source component and an extended component using the JWST images. Hereafter we assume that all point-source components are AGNs, and thus set an absolute upper bound for the AGN population. With this assumption, we calculate a UV LF and measure the maximum ionizing photons produced by AGNs. We adopt a standard $\Lambda$CDM cosmology with $H_0=70$ km s$^{-1}$ Mpc$^{-1}$, $\Omega_\mathrm{m}=0.3$, and $\Omega_\mathrm{\Lambda}=0.7$. All magnitudes are in the AB system. We caution that `AGN' in the literature refers to an AGN with or without its host galaxy. In this paper, `AGN' specifically means the latter (i.e., just a central engine) when it is related to `AGN LF', `AGN component', or `AGN contribution' in situ.

\begin{figure}[h]
\centering
\includegraphics[width=0.73\linewidth]{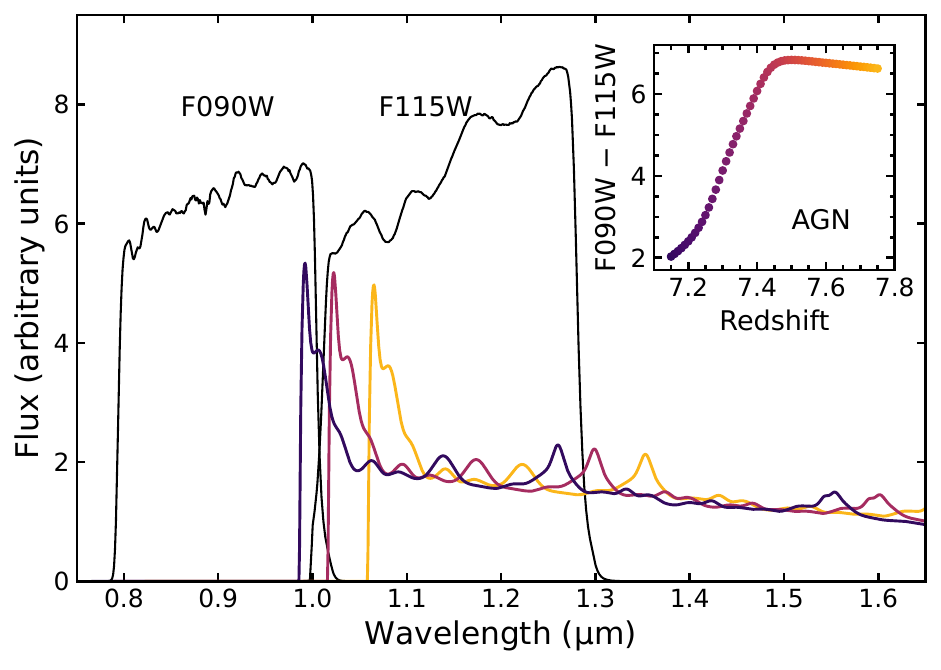}
\caption{\textbf{Color selection of the high-redshift sample.} The three spectra represent a model spectrum of quasars (AGNs) that has been redshifted to $z =$ 7.15, 7.40, and 7.75, respectively (Methods). The line colors of the spectra depend on redshift. The two black filter transmission curves represent JWST F090W and F115W filters. The inset shows how the F090W–F115W color changes with redshift. A galaxy component would slightly increase the F090W–F115W color. We use $\rm F090W-F115W>2$ to efficiently remove low-redshift objects.
\label{fig:color_select}}
\end{figure}

%%Sample selection
We select our high-redshift sample in two steps. We first use the photometric redshifts ($z_{\rm photo}$) of the GOODS fields provided by the JWST Advanced Deep Extragalactic Survey (JADES) program \cite{rieke2023a,eisenstein2023,deugenio2024} to select objects in the required redshift range.We further select relatively bright objects with detection significance $>7\sigma$ in F115W (the primary detection band) using our own combined images (see Methods for details). Our images are slightly deeper than those from the JADES public release because we include images taken by other JWST programs. The depth (5$\sigma$ detection within a $0\farcs24$ circular diameter) of the F115W-band images reach $\sim29.2$ mag in GOODS-N and $\sim30.1$ mag in GOODS-S. The following analyses of the $\rm F090W-F115W$ colors and image decomposition are also based on our images. This first step produces a preliminary sample of 176 objects. In the second step, we implement a color selection criterion $\rm F090W-F115W>2$ (or detection significance $<2\sigma$ in F090W). Figure \ref{fig:color_select} demonstrates our color selection using a model quasar spectrum (Methods).  All objects are visually inspected. Finally, we obtain a sample of 155 objects, consisting of 109 in GOODS-S and 46 in GOODS-N. The areas of the GOODS-S and GOODS-N fields that we use are 67.0 arcmin$^2$ and 56.4 arcmin$^2$, respectively. 

\begin{figure}[h]
\centering
\includegraphics[width=0.95\linewidth]{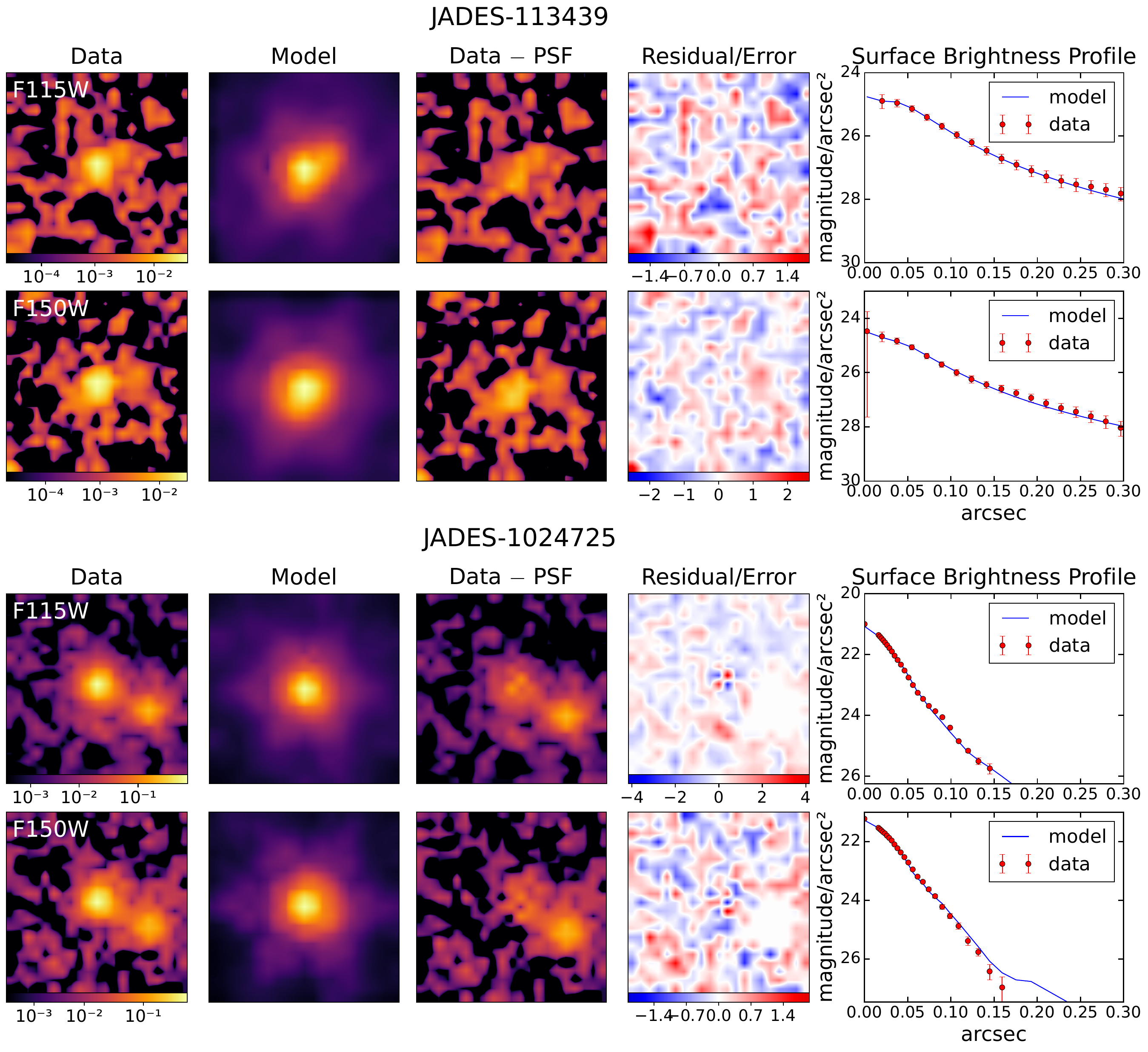}
\caption{\textbf{Two examples of JWST image decomposition in the F115W and F150W bands.} The PSF fractions in the two cases are $\sim 40\%$ and $\sim 70\%$, respectively. The size of each image is 21 pixels (0.63 arc-seconds) on a side, centered on the flux peaks of the targets. The five columns from left to right display the following information: $\left(1\right)$ the observed JWST image; $\left(2\right)$ the best-fit model consisting of a PSF plus a S\'{e}rsic host galaxy; $\left(3\right)$ the PSF-subtracted image; $\left(4\right)$ the fitting residual image divided by the error image; $\left(5\right)$ the surface brightness profile. The vertical bars show the $1\sigma$ errors. The color maps reflect the flux level in the unit of counts per second on a logarithmic scale, as shown by the attached color bars. This image decomposition procedure works well for our sample. 
\label{fig:decomp_images}}
\end{figure}

\begin{figure}[h]
\centering
\includegraphics[width=0.6\linewidth]{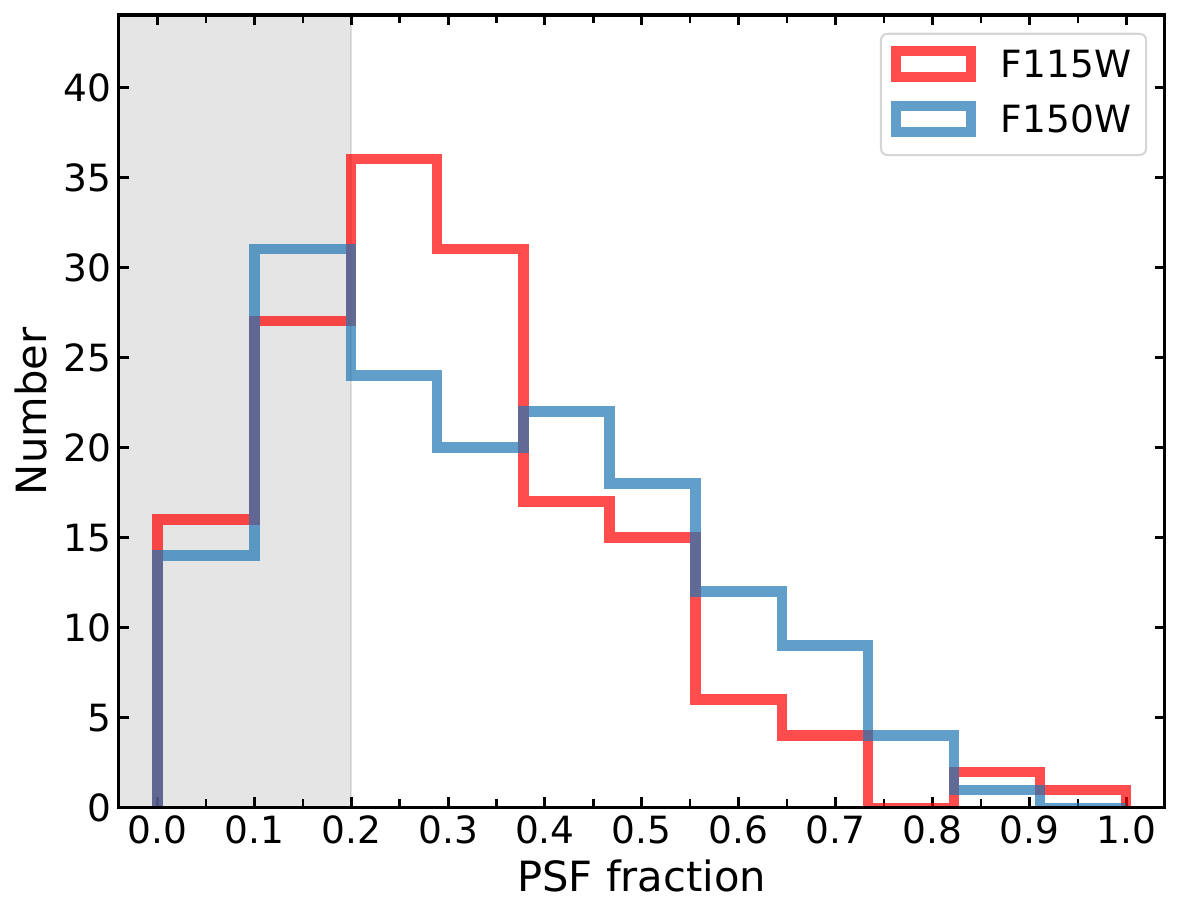}
\caption{\textbf{Distribution of the measured PSF fractions.} The histograms show the PSF fractions for our sample of 155 high-redshift objects in F115W and F150W. The shaded region represents objects with PSF fractions below $20\%$, and they are excluded from our final sample. The two distributions are generally consistent.
\label{fig:agn_fraction}}
\end{figure}

%%image decomposition
We perform the image decomposition \cite{zhuang2023} of the 155 objects using high-resolution JWST images in two rest-frame far-UV bands F115W and F150W (Methods). The decomposition model consists of a point spread function (PSF) and a single S\'{e}rsic profile for AGN and galaxy components, respectively. Figure \ref{fig:decomp_images} illustrates our image decomposition with two examples. 
We define `PSF fraction' of each object as the ratio of the PSF flux to the total flux that represents the maximum AGN fraction. The typical measurement uncertainty of the measured PSF fractions is 5$\%$, which is propagated from the image decomposition. Figure \ref{fig:agn_fraction} shows the distribution of the PSF fractions. We select sources with PSF fractions higher than 20$\%$ in F115W and F150W, i.e., we require that the significance of the PSF detection is higher than $4\sigma$. Objects with lower fractions are completely dominated by galaxies and thus contribute little to the AGN population (Methods). Our final sample contains 100 AGN candidates, consisting of 65 in GOODS-S and 35 in GOODS-N. All image decomposition results for this sample are shown in Supplementary Figure 1. We compute their UV magnitudes at rest-frame 1450\AA~($M_{1450}$) using the decomposed PSF flux. They span a very low luminosity range $-15\leq M_{1450}\leq-20$ mag.

%%LRDs in our sample
We cross match the parent sample of the 155 objects with known LRD candidates in the same regions \cite{kocevski2024,kokorev2024,williams2024}, and find that the sample covers two known LRDs. The other LRDs are either out of the desired redshift range or too faint in the UV band. We further find that our final sample covers one of the two LRDs. The other one is excluded because its PSF fraction is smaller than 20\%. This is because LRDs are primarily identified in the rest-frame optical images, and they can be fainter than our 7$\sigma$ detection limit in F115W and can be extended in the rest-frame far-UV. In the above target selection procedure, we did not use any specific properties of objects other than their $z_{\rm photo}$ and $\rm F090W-F115W$ colors.

%%UV LF: Selection Completeness
We measure the UV LF using the above sample. We first perform simulations to estimate our sample completeness by computing a selection function, i.e., the probability that a galaxy hosting an AGN with a given magnitude $M_{1450}$ at a given redshift $z$ can be selected by our target selection procedure (Methods). We generate two simulated spectral samples for AGN and galaxy components, based on the distribution of the observed PSF fractions in F115W, assuming that the PSF components are all AGNs. The AGN SED model is the same as the quasar model used for our $M_{1450}$ measurement. The galaxy SED is modeled as a power-law continuum, with a slope randomly drawn from a Gaussian distribution based on an empirical relation between UV-continuum slope and UV magnitude of $z \sim 7$ galaxies \citep{bouwens2014a}. The total F090W and F115W magnitudes are then measured from the AGN and galaxy model spectra. We use the total flux of the galaxy without simulating its spatial size, since we also used the total flux for our observational data. With the simulated magnitudes, we add photometric errors for each survey region and calculate the selection function using the standard procedure \cite{matsuoka2018,jiang2016} (Methods). 

\begin{figure}[h]
\centering
\includegraphics[width=0.75\linewidth]{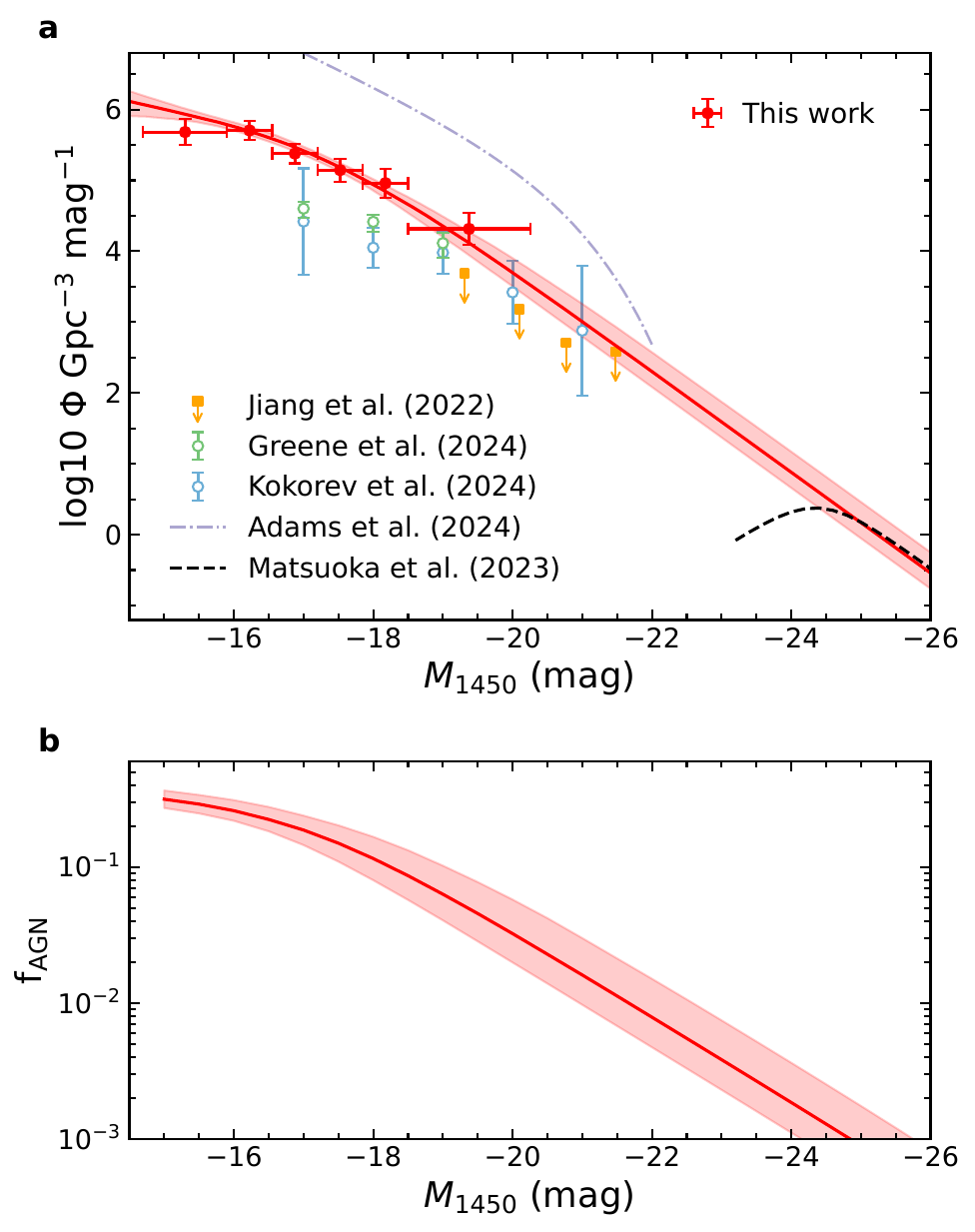}
\caption{\textbf{UV LF and the upper limit of the AGN contribution to reionization.} \textbf{a}, UV LF at redshift $z=7.15-7.75$. The red circles represent our binned LF. Its horizontal bars indicate the luminosity ranges covered by the bins and the vertical bars show the $1\sigma$ errors of the LF. The solid red line with the shaded region represents our best-fit DPL LF with its 1 $\sigma$ region. The light purple dot-dashed line represents a galaxy LF at $z=7.5-8.5$ from ref. \cite{adams2024}. The black dashed line represents the bright quasar LF (QLF) at $z\sim7$ from ref.\cite{matsuoka2023}. The filled orange squares represent the upper limit of QSO LF at $z\sim6.2$ from ref.\cite{jiang2022} based on the observational data before JWST. The light blue open circles represent the LRDs LF at $z=6.5-8.5$ from ref.\cite{kokorev2024}. The light green open circles represent the LF of red AGNs at $z=7-8$ from ref.\cite{greene2024}.
\textbf{b}, Upper limit of the AGN contribution to reionization. The solid red curve with its $\pm1\sigma$ uncertainty region represents the cumulative fraction ($f_{\mathrm{AGN}}$) of the maximum AGN emissivity to the total photon emissivity required to ionize the Universe. They are computed using the above LF. This result excludes the AGN-dominant scenario for cosmic reionization.
\label{fig:UVLF_ion}}
\end{figure}

%UVLF
We compute the binned UV LF using the traditional 1/$V_a$ method \cite{avni1980}. The available cosmic volume for an AGN with $M_{1450}$ and $z$ in a magnitude bin $\Delta M_{1450}$ and a redshift bin $\Delta z$ is given by
$V_a = \int_{\Delta M_{1450}}\int_{\Delta z} p(M_{1450}, z) \frac{dV}{dz} dz dM_{1450}$, where the $p(M_{1450}, z)$ is the selection function derived above. The binned LF is $\Phi(M_{1450}, z) = \sum 1/V_a^i$, where the sum is over all AGNs in the sample. 
We also estimate the uncertainty caused by $z_{\rm photo}$ using Monte Carlo simulations (Methods). The uncertainty of the binned LF is the quadrature sum of the Poisson error and the errors originated from cosmic variance and $z_{\rm photo}$. We ignore the redshift evolution of the LF over the measured redshift range ($7.15\leq z\leq7.75$). The UV magnitude bin size is $\Delta M_{1450}=0.65$, except for the low and high luminosity ends where the sample sizes are small. The calculated binned LF is presented as the red circles in Figure \ref{fig:UVLF_ion}a and listed in Extended Data Table 1.

We further parametrize the LF using a double power-law (DPL) with four parameters, including the faint-end slope $\alpha$, bright-end slope $\beta$, characteristic magnitude $M_{1450}^*$, and normalization factor $\Phi^*$. Since our sample only covers low-luminosity AGNs and the bright end of the LF has been well determined, we fix the bright end slope $\beta$ to be $-2.78$ from the literature \cite{matsuoka2023}. We use a maximum likelihood function \cite{marshall1983} and perform a Markov Chain Monte Carlo (MCMC) analysis \cite{foreman-mackey2013} to determine the best fits of two free parameters $\alpha$ and $M_{1450}^*$. The results are $\alpha=-1.51^{+0.46}_{-0.28}$ and $M_{1450}^*=-17.52^{+0.86}_{-1.16} $ mag. The 1 $\sigma$ uncertainties reported here are derived from the 16th and 84th percentiles of the MCMC parameter distributions. The associated $\Phi^*$ is $10^{-3.5}$ Gpc$^{-3}$ mag$^{-1}$. Because of the high density at the faint end, the LF does not show a distinct `knee'. Therefore, $M_{1450}^*$ has relatively large errors, which has negligible impact on our following calculation. Figure \ref{fig:UVLF_ion}a shows our LF compared with the LFs of quasars, LRDs, and galaxies from the literature \citep{jiang2022,matsuoka2023,adams2024,kokorev2024,greene2024}. Our LF is approximately a few thousand times higher than the extrapolation of the bright quasar LF \citep{matsuoka2023}, tens of times lower than the galaxy LF \citep{adams2024}, and slightly higher than the upper limit of faint AGN LF\citep{jiang2022} based on the observational data before JWST. 

%Contribution to Cosmic Reionization
We follow standard methods \cite{jiang2016,matsuoka2018} to calculate the upper limit of the ionizing photons produced by AGNs using the above LF. To estimate the number of ionizing photons from an AGN, we assume that the escape fraction of ionizing photons is 75\% (ref.\cite{cristiani2016}).
We use the constraint on the EUV slope provided by ref.\cite{Khaire2017} (Methods), and adopt the hardest $\alpha_{\rm EUV}$ of the power-law EUV SED: $f_\nu \propto \nu^{-1.6}$ at $\lambda < 912$\AA, where $\nu$ is frequency. We integrate this SED over an energy range of 1–4 Ryd, and then integrate our LF over a luminosity range from $M_{1450}=-30$ to $-15$ mag to obtain the total ionizing photons from the AGN population. 
The AGN UV radiation is assumed to be isotropic, so the contribution of Type 2 AGNs is included if they are caused by the geometric effect (see more discussion in Methods). The result is compared with the total ionizing photon emissivity per unit comoving volume required to maintain the ionized IGM, which is $\dot{N}_{\text{ion}}(z) = 10^{50.48} (\frac{C}{3}) \times \left( \frac{1+z}{7} \right)^3 \, \text{Mpc}^{-3} \, \text{s}^{-1}$ (ref. \cite{madau1999d}), where $C$ is the IGM clumping factor and we adopt $C=2.55$ at $z=7.5$ (ref.\cite{finlator2012}). In the calculation we have assumed the baryon density of $\Omega_b h^2=0.022$. We define a fraction $f_{\rm AGN}$ as the ratio of the AGN emissivity to the total required photon emissivity. This $f_{\rm AGN}$ ratio represents the maximum AGN contribution to reionization. The result from the above calculations is $f_{\rm AGN} = 32^{+5}_{-4}\%$. Figure \ref{fig:UVLF_ion}b illustrates the cumulative $f_{\rm AGN}$ as a function of luminosity, and shows that low-luminosity AGNs dominate $f_{\rm AGN}$. 
We vary the EUV SED shape and clumping factor $C$ within reasonable ranges, and find that the resultant $f_{\rm AGN}$ values are consistently below $40\%$ (see Methods).

The calculated $f_{\rm AGN} \approx 32\%$ effectively rules out the AGN-dominant scenario for cosmic reionization. We emphasize that this value is the absolute upper bound for two main reasons. One is that the escape fraction of ionizing photons from AGNs has been assumed to be 75\%, including dust extinction. This is a reasonable approximation for luminous quasars\cite{cristiani2016}, but should be smaller in low-luminosity or obscured AGNs \cite{micheva2017,grazian2018,iwata2022}. If it is lower, $f_{\rm AGN}$ would decrease accordingly. 
The second reason is the assumption during the image decomposition procedure. We have assumed that the unresolved PSF component is entirely from an AGN. This ignores the flux from a possible compact galaxy core and overestimates the AGN flux. Some galaxies are very compact and PSF-like even in JWST images\cite{perez-gonzalez2024a,labbe2024}. Considering compact galaxy cores, real AGNs should have lower luminosities and thus move towards the low-luminosity end in the LF, which reduces the overall AGN emissivity. Our result concludes a subdominant role of AGNs for reionizing the Universe, suggesting that star-forming galaxies provide most ionizing photons. It also puts constraints for future reionization models\cite{madau2024}. In the case of $f_{\rm AGN} = 32\%$, the minimum required contribution from galaxies is $\sim 68\%$, which implies an ionizing photon emissivity of log($\dot{N}_{\text{ion}} [\text{Mpc}^{-3}\ \text{s}^{-1}])\geq 50.5$ at $z \sim 7.5$. This estimate is consistent with recent studies \cite{simmonds2024, Choustikov2025}. Future JWST observations will provide further evidence supporting the dominant role of galaxies in cosmic reionization.

\newpage
\noindent \textbf{\Large Methods}

\medskip
\noindent {\bf JWST NIRCam image reduction.} \label{sec:NIRCam_image}
We used raw JWST/NIRCam image files {\tt\string uncal.fits} in the GOODS-S and GOODS-N fields from the MAST archive. The images in the GOODS-S field are from proposal ID 1180, 1210, 1283, 1286, 1895, 1963, 2079, 2198, 2514, 3215, 3990, and 6541, and the images in the GOODS-N field are from proposal ID 1181, 1895, 2514, 2674, and 3577. We reduced the data using the combination of the JWST Calibration Pipeline (v.1.12.5), the scripts for CEERS NIRCam imaging reduction, and our own custom codes. The details of the CEERS pipeline are presented in ref.\cite{bagley2023}. Here we provide a brief description of the steps for our image reduction.

Stage 1 of the JWST Calibration Pipeline performed detector-level corrections, starting from {\tt\string uncal.fits} files and ending with {\tt\string rate.fits} files in units of count per second. Before running the ramp-fitting step in Stage 1 of the pipeline, we identified snowballs, enlarged their footprints, and flagged them as the {\tt\string JUMP\_DET} so that snowballs were properly removed after the ramp-fitting step. ``Wisp'' features from detector B4 of the F150W, F200W, and F210M images were subtracted using the wisp templates provided by the NIRCam team. The horizontal and vertical striping patterns (i.e., 1/f noise) in the images, were subtracted with {\tt\string remstriping.py} in the CEERS team's scripts.

We ran Stage 2 of the JWST Calibration Pipeline with the default parameters, which involve individual image calibrations such as flat-fielding and flux calibration. The output files are {\tt\string cal.fits}. We ran Stage 3 of the JWST Calibration Pipeline to obtain a single mosaic for each filter in each field. The astrometry was calibrated using {\tt\string TweakregStep} with user-provided reference catalogs that were produced as follows. We first generated a WCS reference catalog using the Hubble Legacy Field GOODS-S/GOODS-N F814W image. We then registered  JWST long-wavelength (LW) F277W, F356W, F410M, and F444W images using the F814W WCS. All these LW images were combined to make a master image. Finally, a reference catalog was produced by detecting objects in this master image. 

After running {\tt\string SkyMatchStep} and {\tt\string OutlierDetectionStep}, we subtracted a 2D background for each image following the method described in ref.\cite{bagley2023} and ran {\tt\string ResampleStep} to drizzle and combine images to make one mosaic per filter. We set {\tt\string pixfrac} to 0.8 and pixel scale to $0\farcs03$ for all filters. The {\tt\string output\_shape}, the {\tt\string CRVAL}, and the {\tt\string CRPIX} parameters are the same for each field so that the mosaics of all filters have the same WCS grid. Finally, a 2D background was subtracted from each mosaic using the method described in ref.\cite{bagley2023}.

\medskip
\noindent {\bf High-redshift sample selection.} \label{sec:sample_selection}
We built our sample at $7.15\leq z\leq7.75$. In this redshift range, the strong Ly$\alpha$ break in object spectra results in a very red $\rm F090W-F115W$ color that can efficiently distinguish high-redshift quasars and galaxies from lower-redshift objects. Because the sky coverage of JWST F070W images in the GOODS fields is small and the $\rm F115W-F150W$ color traces the Ly$\alpha$ break at $z\sim10$, the $\rm F090W-F115W$ color is the best choice to investigate high-redshift sources at the epoch of reionization that peaks at $7<z<8$. To demonstrate our color selection, we used a model quasar from ref.\cite{mcgreer2013}, which was also used for our M1450 and UV LF measurement. We applied a typical Ly$\alpha$ damping wing and complete blueward IGM absorption on the model quasar spectrum to simulate the conditions at $z > 7$.

Our sample selection consisted of a $z_{\rm photo}$ selection and a color selection described in the main text. We performed the $z_{\rm photo}$ selection using JADES catalogs. The WCS of our images is slightly different from those used by JADES. In addition, our images are slightly deeper than the JADES images because we included images from programs other than JADES. Therefore, before applying the color selection, we refined the object photometry in F090W and F115W using our combined JWST images.
We used {\tt\string SEP} \cite{barbary2016}, a python version of {\tt\string SExtractor} \cite{bertin1996}, to update the object coordinates using our F115W-band image. We extracted each source in a $21\times21$ pixel$^2$ 2D image cutout centered on its JADES coordinate. We then used the pixel peak coordinate ({\tt xpeak}, {\tt\string ypeak}) as the object coordinate because our work focuses on the central AGN component. With the updated coordinates, we did photometry on the F090W and F115W band images using a circular aperture with $D = 0\farcs24$, about 3.5 times the full width at half maximum (FWHM) of the PSF at F115W. For sources with multiple components in F115W that were not deblended by JADES, we identified their multiple centers and removed those that did not satisfy our selection criteria. All objects are visually inspected.

In the above steps, we did not use morphological information of the objects. Our sample includes both compact and extended sources. The selection criteria mainly rely on the strong Ly$\alpha$ break caused by the strong IGM absorption at high redshift, so our sample is quite complete for galaxies and Type 1 AGNs (completeness correction will be done next) down to the magnitude limit of $7\sigma$ at F115W, the closest band to the LyC wavelength. 
In principle, the selection procedure can also select Type 2 AGNs if their F115W magnitudes satisfy the magnitude limit, because those AGNs at $z>7.15$ should also have a very strong Ly$\alpha$ break. 
Nevertheless, in the following image decomposition, objects with negligible AGN components will not contribute to our calculation.

\medskip
\noindent {\bf Image decomposition.} \label{sec:image_decomp}
Our image decomposition was performed in the F115W and F150W bands. The model consists of a point spread function (PSF) and a single S\'{e}rsic profile to model the point source and host galaxy. The method is developed from the literature \cite{zhuang2023,sun2024,chen2024}, with a multi-band fitting code {\tt\string GalfitM} \cite{2022galfitm}. The PSF was constructed by stacking bright, isolated, and unsaturated point sources in each JWST field, selected based on their small half-light radius and high brightness. For the galaxy component, we restricted the S\'{e}rsic index $n$ to be less than 5, so our fitting tended to model the central component as a PSF component, leading to a conservative upper limit estimation.

In the fitting process, we first used the python codes {\tt\string statmorph} \cite{2019statmorph} and {\tt\string photutils} \cite{larry_bradley_2024} to estimate the position, morphology, and magnitudes for both point source and host galaxy components. The derived values were fed to {\tt\string GalfitM} as input fitting parameters. The PSF position of each source was constrained to be within $\pm1$ pixel around its ({\tt xpeak}, {\tt\string ypeak}) to ensure that the PSF was fitted to the peak of our target. For the multi-component objects, the S\'{e}rsic positions were also limited to be within $\pm1$ pixel around their corresponding peaks. The magnitudes of the point source and host galaxy components were measured using the best-fit PSF model and the PSF-subtracted image, respectively. 
We performed two iterations of the {\tt\string GalfitM} fitting to improve the fitting results. 
After the first iteration, we used the PSF-subtracted images to refine the morphological model of the host galaxy. The derived parameters served as inputs for the second iteration of fitting.

Our image decomposition based on {\tt\string GalfitM} is robust. We adopted a simple two-component model consisting of a PSF and a single S\'{e}rsic profile to minimize possible degeneracy. Recent studies\cite{Zhuang2024} using JWST/NIRCam images have tested AGN+host image decomposition with {\tt\string GalfitM}. By comparing the recovered parameters of mock AGNs with their input values, they found good consistency in the AGN magnitudes, except in cases where the AGN-to-host flux ratio is 0.1, which leads to deviations in the faint end. As described in the main text, we only retained sources with PSF fractions greater than $20\%$, ensuring the reliability of our PSF component measurements. In addition, as discussed above, we adjusted the fitting parameters to reduce the risk of the minimization algorithm in {\tt\string GalfitM} converging to local minima. We also visually inspected all fitted objects to ensure the decomposition results are physically reasonable.

We estimated the uncertainty caused by the selection of point sources in PSF construction. 
We reconstructed PSFs for target sources based on different selection criteria, consisting of the brightest, faintest, nearest, and farthest five point sources. We then performed image decomposition on the target sources using these varied PSFs and computed $M_{1450}$. Comparing these results with those obtained using the standard PSF, we found that the uncertainties introduced by point source selection are mostly within the measurement uncertainty of individual $M_{1450}$. Furthermore, as shown in the next section, the uncertainty in $M_{1450}$ has a negligible impact on our UV LF and $f_{\rm AGN}$, indicating that the uncertainties from PSF point source selection are also negligible.

\medskip
\noindent {\bf UV LF.} \label{sec:UVLF}
To construct the UV LF, we first measured $M_{1450}$ of our AGN candidates using their magnitudes in F115W and F150W. We derived $k$-corrections in the two bands as follows. For each band, we generated a grid of simulated quasars in the [$M_{1450}$, $z$] space with $\Delta M_{1450} = 0.25$, $\Delta z = 0.05$. The quasar spectral energy distributions (SEDs) from ref.\cite{mcgreer2013} were designed to reproduce the colors of $\sim 60,000$ quasars. Assuming that the quasar SEDs do not evolve with redshift, we extended the quasar SED model to redshifts $z>7$. We further assumed a typical Lyman-alpha damping wing absorption for the spectra following observational results\cite{Banados2018,wang2020}. The blue side of the damping wing was completely absorbed by the IGM. We used the simulated spectra to calculate the $k$-corrections as a function of redshift and luminosity. The $k$-corrections of the F115W and F150W magnitudes for an individual AGN were then interpolated from their grids, respectively. We calculated two values of $M_{1450}$ based on the F115W and F150W magnitudes and adopted their average value as the final AGN $M_{1450}$ for this AGN. 

We note that the above SED model was derived from luminous quasars. On one hand, it ensures the SED is dominated by the AGN radiation without much host galaxy contamination. On the other hand, these quasars are much more luminous than the objects in our sample, and may have different SEDs that could affect our results. We found that the potential effect is negligible. For the calculation of $M_{1450}$, we compared the above result with an extreme case in which an SED model is a pure power-law continuum without emission lines, and the difference is smaller than 5\%. We also used this SED model for the color selection previously, but our color selection criteria mainly rely on the IGM absorption feature. Therefore, these calculations are insensitive to specific SED shapes.

We also applied an alternative method to estimate $M_{1450}$. For each object, we calculated its flux at rest-frame 1285\AA~and 1835\AA~using the F115W and F150W magnitudes, where $k$-corrections were derived as previously described. We then fitted a power-law continuum to the flux and computed $M_{1450}$ from the continuum. The resultant $M_{1450}$ values agree well with those obtained from the first method. For simplicity, we used the average $M_{1450}$ values from the first method for the following analyses.

We then estimated our sample completeness by computing a selection function as described in the main text. With the simulated F090W and F115W magnitudes, we added photometric errors using the standard procedure \cite{matsuoka2018,jiang2016}. Following the magnitude versus error diagram of the observed objects in the GOODS fields, we drew a large sample of simulated objects whose distribution in the magnitude versus error diagram is the same as that of the real objects. Note that the detection depth in the GOODS fields are not uniform because we combined many JWST images to take advantage of the available observations. Our procedure can properly handle this. The areas of GOODS-S and GOODS-N that we used are 67.0 arcmin$^2$ and 56.4 arcmin$^2$, respectively. The whole GOODS-N field and 25 arcmin$^2$ in GOODS-S field reach $\sim29.2$ mag ($5\sigma$ detection in a $D = 0\farcs24$ circular aperture) in F115W. The region of the other 42 arcmin$^2$ in GOODS-S is deeper ($\sim30.1$ mag). Finally, we calculated the fraction of sources in each $M_{1450}$ and $z$ bin that satisfied our color selection criteria. The UV LF was then constructed considering the sample completeness. 

We evaluated the impact of $M_{1450}$ measurement uncertainties on the UV LF and $f_{\rm AGN}$ using a Monte Carlo (MC) simulation. We generated 1000 MC mock samples, where each $M_{1450}$ was randomly drawn from a Gaussian distribution centered on its measured value, with a standard deviation equal to its uncertainty. 
The UV LF calculation followed the same procedure as for the real sample. To isolate the effect of $M_{1450}$ uncertainties, we fixed the bright-end slope and disregarded errors in the literature parameters. As a result, the propagated uncertainty in $f_{\rm AGN}$ is only $0.5\%$, and the impact on the UV LF is also negligible. Therefore, we did not consider $M_{1450}$ measurement uncertainties in our main analysis.

We also estimated the uncertainties caused by the $z_{\rm photo}$ measurements using a method similar to those in the literature\cite{kokorev2024,Marchesini2009}. We first constructed an enlarged sample based on the same source selection criteria described in the main text but with a broader redshift range of $6.5 \leq z \leq 8.5$. We then performed image decomposition for each object. Using quantiles of the redshift probability distribution $P(z)$ released by JADES, we reconstructed the individual $P(z)$. We generated 3000 MC mock samples using $P(z)$ for the enlarged sample. For each mock sample, we selected sources within our target redshift range of $7.15 \leq z \leq 7.75$, and calculated the LFs using the same procedure as for the real sample.
Note that since we also applied the color selection criteria to the enlarged sample, which effectively excluded sources with actual redshifts below 7.15, the impact of $z_{\rm photo}$ uncertainties was reduced. 
The results show that the typical uncertainty in the binned LF caused by $z_{\rm photo}$ measurements is $\log \Phi[\text{Gpc}^{-3}\ \text{mag}^{-1}] \sim0.06$. Individual uncertainty has been included in the error bar of each LF bin in the main results.
The uncertainty of $f_{\rm AGN}$ caused by $z_{\rm photo}$ is only $2\%$, which has a negligible impact on our main conclusion. Therefore, we did not include the $z_{\rm photo}$ uncertainties in our main parametrized LF and $f_{\rm AGN}$.

In order to assess the impact of objects with lower AGN fractions, we also constructed a sample with PSF fractions higher than 10$\%$. This larger sample contains 132 AGN candidates, consisting of 91 in GOODS-S and 41 in GOODS-N. The objects with PSF fractions of $10-20\%$ are mostly distributed in a low luminosity range of $M_{1450}=-17\sim-14$. We applied the above methods to this sample and calculated the $f_{\rm AGN}$. Under the same assumptions, the resultant $f_{\rm AGN} = 31^{+4}_{-3}\%$, which is well consistent with our standard result. 
The $f_{\rm AGN}$ insignificantly decreased by $0.5\%$, which is attributed to a minor change in the faint-end slope of the LF caused by the inclusion of more data points in the low luminosity range.
The AGN contribution does not increase because the faint-end AGN LF has a slope with an expectation that there are more fainter AGNs. If the density of the newly added (very faint) AGNs does not exceed this expectation, the faint-end LF does not increase, and the AGN contribution does not increase.
This result supports the robustness of selecting sources with PSF fractions larger than 20$\%$. Sources with lower PSF fractions are dominated by galaxies with negligible contributions from their AGN components.

\medskip
\noindent {\bf AGN Contribution to Cosmic Reionization.} \label{sec:fAGN}
When we calculated the total AGN LyC photons in the main text, we have assumed that these AGNs are Type 1 AGNs, and their UV radiation is isotropic. The AGN unification model suggests that AGNs appear as Type 1 and Type 2 due to the geometric effect or different orientation angles. LyC photons from certain angles or directions can leak out of the AGN system, while LyC photons in the other directions are absorbed and do not contribute to reionization. Therefore, the assumption of the isotropic radiation above accounts for both Type 1 and Type 2 AGNs.
While the AGN unification model is well established and widely accepted, there are other AGN models. We briefly investigate the impact of some of these models on our result. For example, dense interstellar medium (ISM) in a host galaxy can further obscure its AGN \citep{AlonsoT2024,Andonie2024}. Observed Type 2 AGNs may have a larger torus covering factor and a lower escape fraction compared to Type 1 AGNs \citep{RamosA2011}. In such cases, the escaped ionizing photon flux would be further reduced, which does not affect our upper limit. The effect of randomly oriented galaxy disks should have been included in our isotropic integral. More complex models are beyond the scope of this work.

AGNs have duty cycles. The required ionizing photon emissivity is used to balance the recombination at $z\sim7.5$ (or at a given moment), so we directly used the calculated LyC photon emissivity from AGNs without considering AGN photons in the past. The photons produced in the past are used to balance the recombination in the past. One can compare the accumulated total ionizing photons from AGNs and the accumulated ionizing photons consumed by recombination in a period of time. Here we assume that they do not change much in a short period of time.

Based on our standard calculation of the total ionizing photon emissivity described in the main text, we evaluated the effects of different parameter choices on our $f_{\rm AGN}$ result. It is still infeasible to construct the EUV SED for the high-redshift AGN sample. We considered the constraint on the EUV slope from the literature\cite{Khaire2017}, which used opacity measurements of the He II Ly$\alpha$ forest at $2.5 < z < 3.2$ from ref.\cite{Worseck2016}, and concluded that the EUV slope $\alpha_{\rm EUV}$ must lie in the range of $-2<\alpha_{\rm EUV}<-1.6$, with a preferred value of $-1.8$. This range of slopes resulted in a range of $f_{\rm AGN} = 27\%-32\%$, with $f_{\rm AGN} = 29\%$ when $\alpha_{\rm EUV} = -1.8$. Given the recent discoveries of high-redshift AGNs with relatively low black hole masses, we adopted the hardest EUV slope $\alpha_{\rm EUV} = -1.6$ in our main calculation. This choice is also conservative for estimating the upper limit of $f_{\rm AGN}$. We also varied the value of the IGM clumping factor C. Simulations have suggested C$\sim2-5$ (ref.\cite{madau2024,finlator2012,Shull2012,McQuinn2011}), corresponding to $f_{\rm AGN}= 16\%-40\%$. 
In the above calculations, we have assumed a very high LyC escape fraction (75\%) from AGNs. In reality, it is lower, as we mentioned in the main text. In addition, we have assumed that the PSF component is completely dominated by the AGN, which neglects any stellar contribution. Therefore, the real AGN contribution should be considerably smaller than the estimated values above.

\subsection*{Data Availability}
All JWST data used in this paper are publicly available in the MAST archive \citep{DOI}.
The wisp templates used in the image reduction are provided by the NIRCam team: \url{https://stsci.app.box.com/s/1bymvf1lkrqbdn9rnkluzqk30e8o2bne}.

\subsection*{Code Availability}
Data were reduced using publicly available data reduction pipelines\cite{bagley2023}: \url{https://github.com/ceers/ceers-nircam}. Public tools were used for data analysis: {\tt\string SEP} \cite{barbary2016}, {\tt\string GalfitM} \cite{2022galfitm}, {\tt\string statmorph} \cite{2019statmorph}, {\tt\string photutils} \cite{larry_bradley_2024}.

\subsection*{Acknowlegements}
D.J., L.J., S.S., W.L. and S.F. acknowledge support from the National Science Foundation of China (12225301).
This work is based on observations made with the NASA/ESA/CSA James Webb Space Telescope. The data were obtained from the Mikulski Archive for Space Telescopes at the Space Telescope Science Institute, which is operated by the Association of Universities for Research in Astronomy, Inc., under NASA contract NAS 5-03127 for JWST. These observations are associated with program 1180, 1210, 1283, 1286, 1895, 1963, 2079, 2198, 2514, 3215, 3990, 6541, 1181, 1895, 2514, 2674, and 3577.
The authors acknowledge the JADES team led by PIs Daniel Eisenstein, for developing their observing program with a zero-exclusive-access period.

\subsection*{Author Contributions Statement}
D.J. analyzed the data, performed the calculations, and wrote the manuscript. L.J. designed program and wrote the manuscript. S.S. helped with the image decomposition. W.L. performed the JWST NIRCam image data reduction. S.F. helped with the target selection. S.S., W.L., and S.F. all contributed to the relevant sections of the manuscript.

\subsection*{Competing Interests Statement} 
The authors declare that they have no competing interests.

\clearpage
%% Figures
\section*{Figure}

\end{document}